\documentclass[epj]{svjour}
\usepackage{graphics}
\usepackage{epsfig}
\begin{document}
\title{How to Make Sense of the Jet Correlations Results at RHIC?}
\subtitle{}
\author{Jiangyong Jia\inst{1} \inst{2}% etc
}                     % Do not remove
\offprints{jjia@bnl.gov}          % Insert a name or remove this line
\institute{Chemistry Department, Stony Brook
University, Stony Brook, NY 11794, USA\\
\and Physics Department, Brookhaven National Laboratory, Upton, NY
11796, USA }
\date{\today}
%, which takes into account importatrigger contributions from medium response

 \abstract{ We review the di-hadron correlation results
from RHIC. A consistent physical picture was constructed based on
the correlation landscape in $p_T$, $\Delta\phi$, $\Delta\eta$ and
particle species. We show that the data are consistent with
competition between fragmentation of survived jets and response of
the medium to quenched jets. At intermediate $p_T$ where the medium
response are important, a large fraction of trigger hadrons do not
come from jet fragmentation. We argue that these hadrons can
strongly influence the interpretation of the low $p_T$ correlation
data. We demonstrate this point through a simple geometrical jet
absorption model simulation. The model shows that the correlation
between medium response hadrons dominates the pair yield and mimics
the double hump structure of the away-side $\Delta\phi$
distribution at low $p_T$. This correlation was also shown to lead
to complications in interpreting the results on reaction plane
dependence and three particle correlations. Finally, we briefly
discuss several experimental issues which are important for proper
interpretations of the experimental data.
\PACS{      {25.75.}{-q} } % end of PACS codes
} %end of abstract
\maketitle
\section{Introduction}
\label{intro}

Single jets and back-to-back di-jet pairs are important tools for
studying the properties of the dense matter created in relativistic
heavy-ion collisions. Due to difficulties of full jet
reconstruction at RHIC, they are accessed through leading hadron
spectra and di-hadron correlation. The primary handle for spectra
analysis is the hadron $p_T$, whereas multiple handles can be used
in di-hadron correlation analysis: the momentum of trigger
($p_T^A$), momentum of partners ($p_T^B$), $\Delta\phi$ and
$\Delta\eta$. The possibility of varying all these variables leads
to large amount of experimental data.

\begin{figure}
\resizebox{1\linewidth}{!}{%
  \includegraphics{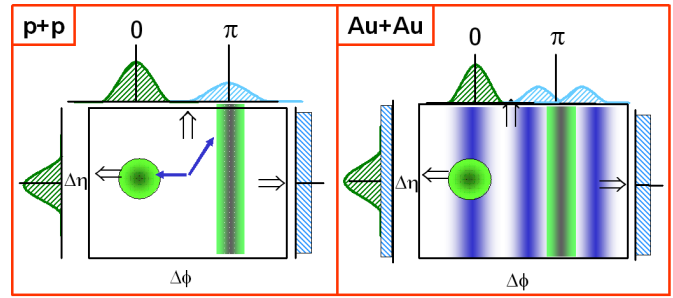}
}
\caption{\label{fig:1} A schematic illustration of the jet-induced di-hadron correlation signal in $\Delta\phi$ and $\Delta\eta$ and
associated projections in $\Delta\phi$ or $\Delta\eta$ for (left) $p+p$ and (right) central Au+Au collisions.
The medium response components in Au+Au is illustrated by fuzzy blue band.}
\end{figure}

A schematic illustration of the di-hadron correlation signal for
unmodified jets (such as in p+p collisions) is presented in
Fig.~\ref{fig:1}a. The signal appears as a narrow peak at
$(\Delta\phi,\Delta\eta)\sim(0,0)$ and a broad peak at
$\Delta\phi\sim\pi$ which is flat in $\Delta\eta$ up to
$|\Delta\eta|<2$. The former corresponds to pairs from the same
jet, while the later corresponds to pairs from the away-side jet.
The elongation of the away-side pairs in $\Delta\eta$ is due to
longitudinal momentum imbalance between the two original partons
that undergo hard-scatter process.

In Au+Au collisions, jets are modified by the dense medium. In
addition to the jet fragmentation component coming mainly from
those jets that do not interact with the medium (due to surface
emission or punch-through)~\cite{Adams:2006yt,Adare:2007vu},
extensive experimental
studies~\cite{Adams:2005ph,Adler:2005ee,Putschke:2007mi,Adare:2008cq}
have revealed additional medium response components as illustrated
in Fig.~\ref{fig:1}b. These medium response components appear as
three distributions that peaks at different $\Delta\phi$ locations:
$\Delta\phi\sim0,\pi\pm1.1$, but are flat in $\Delta\eta$. The
current focus of the field is to understand the interplay of the
contributions from jet fragmentation and medium response.

Beyond this qualitative picture of the jet-medium interaction, the
progress so far is limited in our understanding of the fundamental
mechanisms in terms of the plasma properties. We do not yet fully
understand the jet quenching, let alone the medium response
mechanism. In a sense there are too much experimental information:
new measurements are continuously being made, but the progress of
integrating them into fundamental theoretical framework has been
slow.

The reason for current impasse is two fold. Experimentally, there
are assumptions made in the extraction of jet signal, such as the
two-source model and ZYAM assumptions in background
subtraction~\cite{Adams:2005ph,Adler:2005ee}. There are also
questions on how to deal with non-flow effects and event-by-event
fluctuation in background subtraction. On the theory side, the
medium response is intrinsically non-perturbative and difficult to
model. Most calculations assume that there is a scale that
separates jet energy loss and medium
response~\cite{Casalderrey-Solana:2006sq,Renk:2006pk}. The modeling
of energy transport to medium response is done in an ad hoc
fashion, which necessary introduce a scale dependence. The ADS/CFT
based approaches can describe both jet energy loss and Mach-cone
like medium response in a single
framework~\cite{Gubser:2008as,Chesler:2007an,Friess:2006fk}.
However it is not clear that these results can be extrapolated to
QCD. The next improvement is to combine the calculation of jet
quenching and medium response with the 3-D hydrodynamics
framework~\cite{techqm}.

How can we as experimentalist help with current situation? Given
the vast amount of data and the experimental difficulties mentioned
earlier, we feel it is imperative to re-exam all the data and
construct a global and self-consistent physical picture. A
convincing picture should be able to connect results for high and
low $p_T$, near- and away-side, and encompass different particle
species and collision energies.

In the process of understanding underlying mechanisms for
modification of jet correlation, we should not separate them from
the mechanisms that are important for single particle spectra and
flow measurements, such as hydrodynamic expansion and
recombination. In fact, these bulk mechanisms should play a very
important role in shaping the jet correlation pattern. The physical
picture implied by correlation and bulk measurements should be
consistent with each other. The second task of this manuscript is
to exam the origin of triggers and the sources of correlated hadron
pairs, such that we understand how the bulk production mechanism
and medium response influence the interpretation of the correlation
data.

In the end of the manuscript, we also discuss several experimental
issues which are important for proper interpretations of the low
$p_T$ correlation data.

\section{Correlation Landscape}

\begin{figure*}[!t]
\begin{center}
\resizebox{0.9\textwidth}{!}{%
  \includegraphics{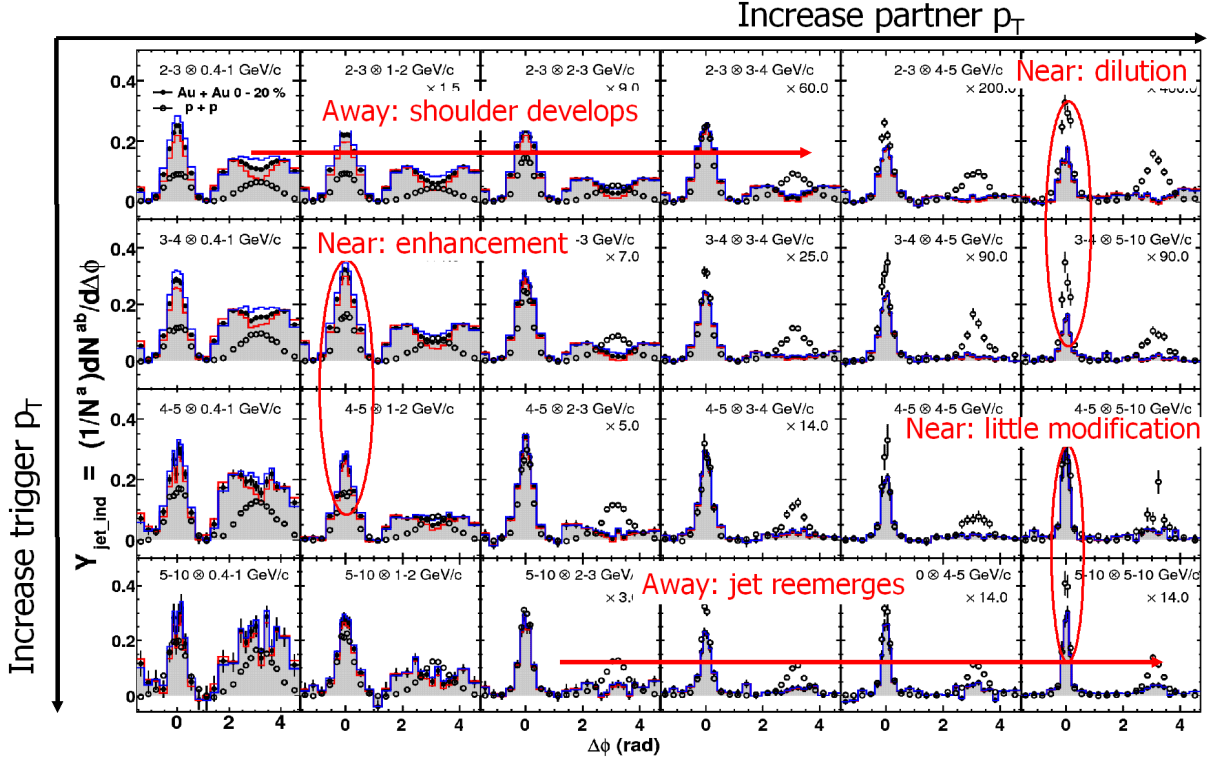}
} \caption{\label{fig:2} The $\Delta\phi$ distribution in fine bin
of trigger and partner $p_T$ in  $p+p$ (open symbols) 0-20\%
central Au+Au (solid symbols) collisions, and thin lines indicates
the systematic uncertainties. Several important features are
indicated by the thick red lines and red circles.}
\end{center}
\end{figure*}

The primary goal of the correlation analysis is to understand the
interactions between jet and medium, i.e. jet quenching and the
medium response. We now know that the jet quenching is responsible
for the large suppression of the jet pairs at high $p_T$. The
medium response is responsible for the enhancement of hadron pairs
at low $p_T$. However most measurements typically focus on a
particular momentum range or particular $\Delta\phi$ range. The
methods and systematical errors associated with these measurements
are not always the same, which makes the comparison of these
results quite difficult. In a recent paper~\cite{Adare:2008cq},
PHEINX collaboration publish a detailed survey of the di-hadron
correlation in broad ranges of trigger and partner $p_T$, which
allows a systematic comparison of large amount data in a single
analysis. Fig.~\ref{fig:2} shows the summary of the correlation
landscape in $p_T^A$, $p_T^B$ and $\Delta\phi$. Many features can
be identified at both the near- and away-side. They reflect in
detail how the jet quenching and medium response vary with momentum
and $\Delta\phi$.

\begin{figure}[h]
\resizebox{1\linewidth}{!}{%
  \includegraphics{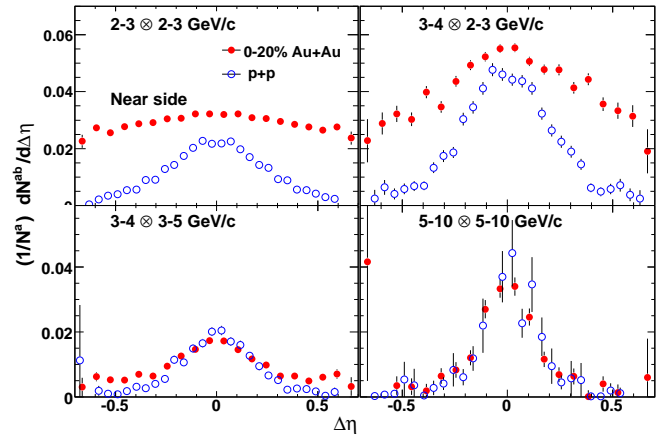}
}
\caption{\label{fig:b2} Per-trigger yield versus
$\Delta\eta$ for $p+p$ (open symbols) and 0-20\% central Au+Au (filled
symbols) collisions. Results are shown for four
$p_{T}^{A}\otimes p_{T}^{\rm B}$ selections as indicated.}
\end{figure}

A systematic approach to summarize these features can be done by
dividing the $\Delta\phi$ distributions into three regions, an
away-side head region ($|\Delta\phi-\pi|<\pi/6$), an away-side
shoulder region ($\pi/6<|\Delta\phi-\pi|<\pi/3$), and a near-side
region ($\Delta\phi<\pi/3$), and then project them in the
$\Delta\eta$ direction. The division of the away-side into the head
and the shoulder regions facilitates the separation of the jet
fragmentation contribution and medium response (the cone) at the
away-side. The projection of near-side pairs in $\Delta\eta$ help
us to separate the jet fragmentation contribution and medium
response (the ridge) at the near-side. The $\Delta\eta$ projections
for near-side pairs are shown in Fig.~\ref{fig:b2} for several
representative $p_T^A\otimes p_T^B$ bins.

The information in Fig.~\ref{fig:2} and Fig.~\ref{fig:b2} can be
captured by a set of shape and yield variables. Jet shape variables
include the ratio of head yield to shoulder yield $R_{\rm HS}$,
mach cone location $D$ or near-side width~\cite{Adare:2007vu}. Jet
yield variables include per-trigger yield (PTY) and $I_{AA}$ in the
three $\Delta\phi$ ranges. A careful analysis of these variables as
functions of $p_T^A$ and $p_T^B$ shows that all the features in
Fig.~\ref{fig:2} and Fig.~\ref{fig:b2} can be explained by the
combined $p_T$ dependence of the jet fragmentation and medium
response components at both the near- and away-side. The jet
fragmentation contribution dominates at high $p_T$ and the medium
response dominates at low $p_T$. The transitional region where the
two sources roughly equal to each other can be approximated by
$p_T^A+p_T^B\sim6-8$ GeV/c.

At the high $p_T$ region where the jet fragmentation dominates, we
normally use the $I_{AA}$ to quantify the away-side suppression. It
was observed that high $p_T$ $I_{AA}$ is similar to single particle
suppression, $R_{AA}$~\cite{Adams:2006yt,Adare:2008cq}. This
observation is surprising at first sight, given that the away-side
jets suffer on average more energy loss than that for inclusive
jets. However, we should realize that the suppression level depends
not only on the energy loss, but also on the jet spectral shape. As
argued in~\cite{Jia:2007qi}, the away-side hadron spectra
associated with a high $p_T$ trigger is much flatter than that of
the inclusive hadrons. A flatter spectrum needs more energy loss to
achieve a given level of suppression. Even though $I_{AA}\sim
R_{AA}$, the away-side jets actually suffer on average 50\% more
energy loss than that for inclusive jets.

At the low $p_T$ region, medium response contributions, i.e.
\emph{the cone} on the away-side and \emph{the ridge} on the
near-side, play an important role in shaping the $\Delta\phi$ and
$\Delta\eta$ distribution. A careful comparison between the ridge
and the cone, after suppressing the jet fragmentation contribution,
indicates that their properties are very similar. Both have similar
spectral slopes, which are harder than inclusive hadrons but are
softer than fragmentation hadrons~\cite{afranzqm}. Both have
similar particle composition in terms of baryon to meson
ratios~\cite{Afanasiev:2007wi,Bielcikova:2008ad}: the ratios are
bigger than that expected for fragmentation and are close to that
for the inclusive spectra (with large uncertainty). These
similarities suggest that mechanisms for the ridge and the cone
could be related. They also suggests that the bulk physics,
hydrodynamical flow and parton recombination, should play an
important role in this $p_T$ region.

Most previous low $p_T$ correlation analyses choose to use
$I_{AA}$, the ratio of per-trigger yield in Au+Au collisions to
that in $p+p$ collision, to characterize the medium response. The
rationale being that since each trigger tags one jet, the
per-trigger yield is a good approximation to the per-jet yield
(similar to high $p_T$ correlation analysis), thus $I_{AA}$
measures the modification of a jet in Au+Au relative to that in
$p+p$. As we shall show in the next section this assumption is not
true at $p_T<5$ GeV/c, and an alternative observable is needed for
describing the medium response.

\section{Origin of Triggers: Jet fragmentation? Bulk? or Medium Response?}
\begin{figure}
\begin{center}
\resizebox{0.47\linewidth}{!}{%
  \includegraphics{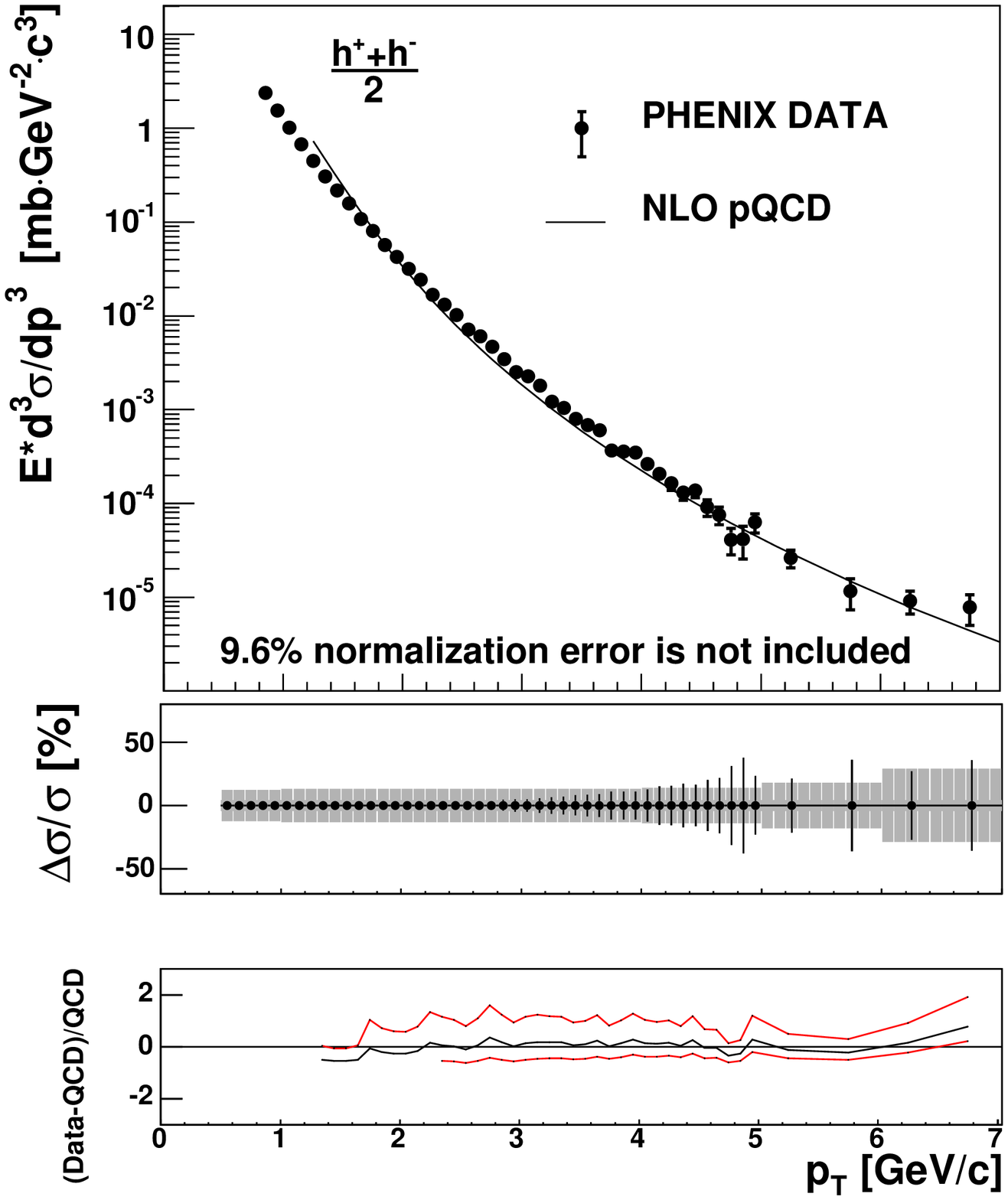}
}
\resizebox{0.47\linewidth}{!}{%
  \includegraphics{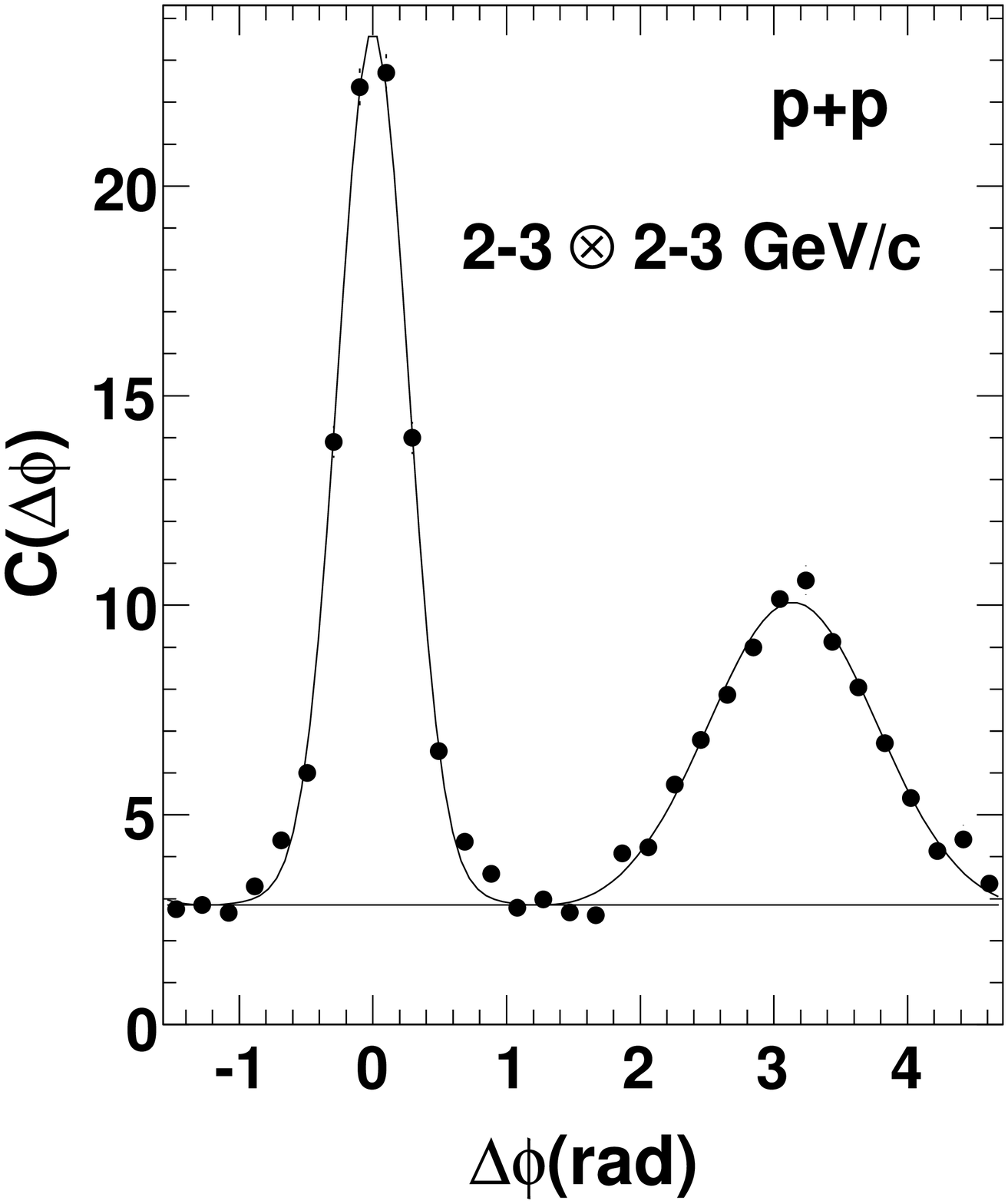}
}
\caption{\label{fig:3} (left) Inclusive charged hadron spectra from $p+p$ collisions compared with pQCD calculation down to 1.5 GeV/c~\cite{Adler:2005in}.
(right) Low $p_T$ di-hadron correlation signal for charged hadrons in $p+p$. Both are for $\sqrt{s}=200$ GeV.}
\end{center}
\end{figure}

In $p+p$ collisions at RHIC energies, data from the single particle
spectra and di-hadron correlation suggest that most hadrons above 2
GeV/c originate from jet fragmentation. This is attested by the
fact the particle spectra are well described by the pQCD
calculation~\cite{Adler:2003pb,Adler:2005in} and the correlation
function show strong jet-like signals down to very low
$p_T$~\cite{Adler:2005ad,Adare:2008cq} as shown by
Fig.~\ref{fig:3}.

\begin{figure*}[!ht]
\begin{center}
\resizebox{0.7\linewidth}{!}{%
  \includegraphics{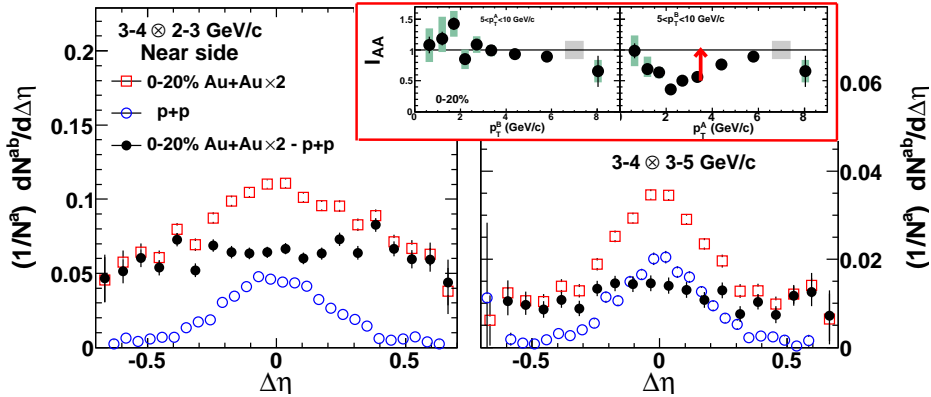}
} \caption{\label{fig:4}  Per-trigger yield $\Delta\eta$
distribution for 3-4 GeV/$c$ triggers and two parter $p_T$
selections. The ridge distribution (solid circles) is estimated by
subtracting the Au+Au distribution corrected by dilution effect
(open squares) minus the p+p (open circles). The dilution
correction ($\times2$) is indicated by the red arrow in the
inserted panel (see text for explanation).}
\end{center}
\end{figure*}

The situation is very different in heavy ion collisions. At $p_T<5$
GeV/c where medium response is important, the inclusive hadron
yield is dominated by non-perturbative bulk production mechanisms
such as hydrodynamic flow and recombination. The experimental
evidences include: a particle composition that is strongly strongly
modified relative to that in $p+p$ collisions~\cite{Adler:2003cb},
a $R_{AA}$ that peaks around $p_T\sim3$ GeV/c at a level much
bigger than what is suggested by jet quenching~\cite{Adler:2003au},
and a hadron elliptic flow that reaches maximum in this momentum
region at a level much bigger than what one expect from the jet
quenching~\cite{Adams:2004wz}. These results imply that the origins
of the hadrons at $p_T<5$ GeV/c in Au+Au collisions are very
different from those in $p+p$. Thus per-trigger yield and $I_{AA}$
are not good observables for the medium response since the triggers
are also modified. According to hydro + coalescence model, a large
fraction of the triggers at $p_T<5$ GeV/c may come from
thermal-thermal or thermal-shower recombination. The former has no
jet correlation, while the latter can retain some jet correlation
which can be modified by the flow~\cite{Fries:2004hd}.

If most hadrons do not originated from jet fragmentation, and have
either no correlation or reduced correlation strength, they should
lead to a dilution of per-trigger yield in Au+Au collisions
relative to $p+p$. In fact PHENIX data indicate the effect of the
dilution on the per-trigger yield~\cite{Jia:2008dp}. It is best
illustrated with the near-side $\Delta\eta$ correlation as shown in
Fig.~\ref{fig:4}~\footnote{The data are from Fig.~\ref{fig:b2}.}.
We estimate dilution factor ($\sim2$) for 3-4 GeV/$c$ triggers
based on their correlations with 5-10 GeV/$c$ hadrons as shown by
the inserted panel: requiring 5-10 GeV/$c$ hadrons ensures the
pairs are dominated by the jet fragmentation (left panel of the
insert), thus deviation of $I_{AA}$ from one for soft triggers
reflects the level of dilution (the red arrow). Once the dilution
factor is corrected, we subtract out the jet fragmentation
contribution and obtain the ridge distribution (black filled
circles). The estimated ridge contribution is approximately flat,
consistent with experimental data at large
$\Delta\eta$~\cite{Putschke:2007mi}. However, this dilution effect
was not observed in some STAR
analyses~\cite{Putschke:2007mi,Bielcikova:2007mb}, which showed
that the PTY$_{AA}$ subtracted by the estimated ridge equals
PTY$_{pp}$ before any correction for dilution effect.

Many previous analyses assume that most triggers originated from
jet fragmentation. This naturally leads to different
interpretations for the medium responses between the near- and
away-side. The near-side pairs were thought to be surface biased
and should have smaller modifications, while the away-side pairs
traverse on average a longer path, thus should be strongly
modified. However, this interpretation is not correct if most
triggers come from the bulk or medium response. In this case, the
triggers are not surfaced biased but are emitted from the whole
volume of the overlap region. We shall elaborate on this point in
the next Section.

The interpretation of the PTY as a function of angle with respect
to the reaction plane, $\phi-\Phi_{RP}$, should also be modified if
non-jet triggers dominate. This is because the variation of trigger
yield with $\phi-\Phi_{RP}$ (characterized by $v_2$) is much larger
than what is expected from fragmentation of survived jets. Thus the
dilution of PTY could also depend on $\phi-\Phi_{RP}$.

\begin{figure}[h]
\begin{center}
\epsfig{file=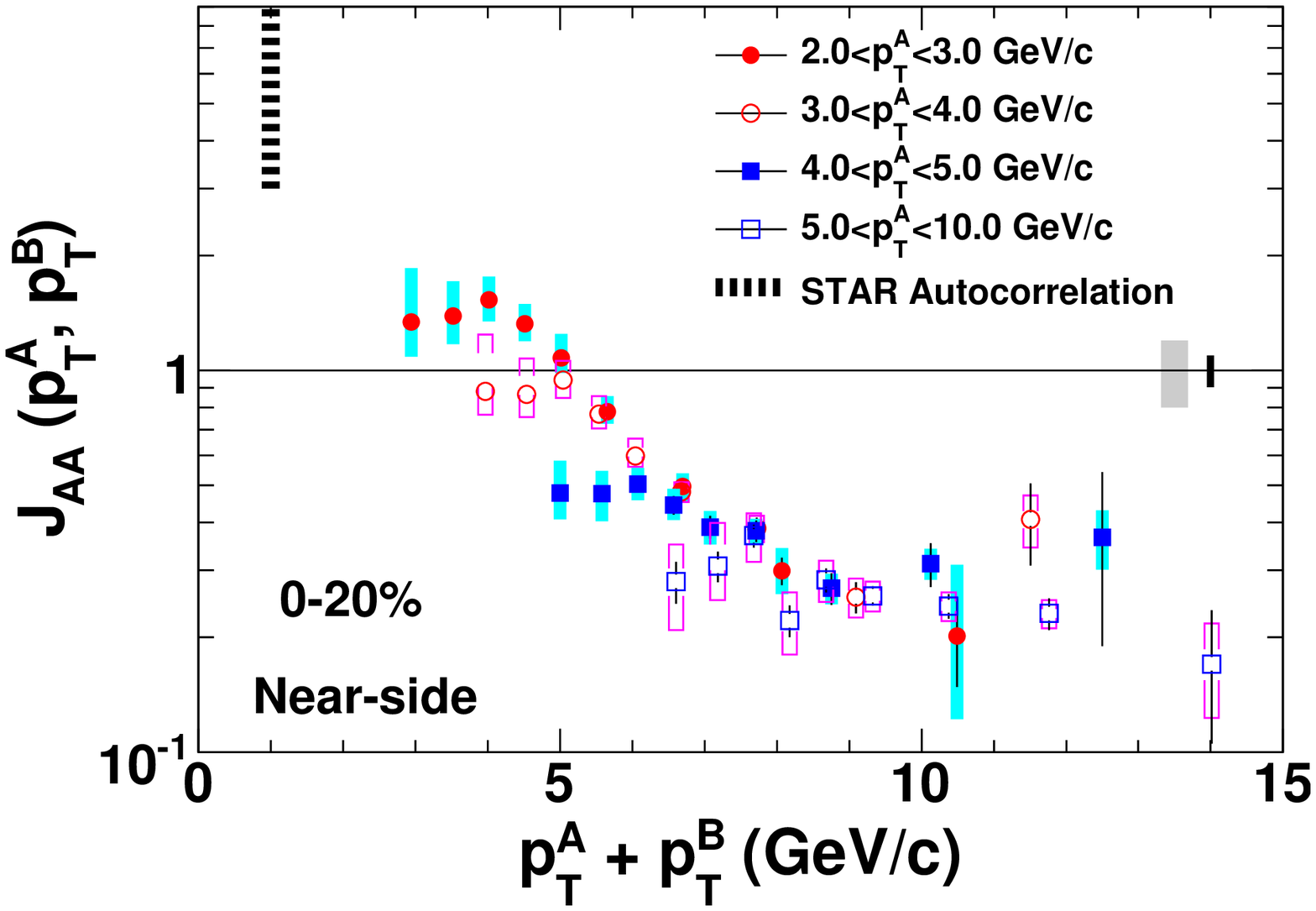,width=0.85\columnwidth}
\epsfig{file=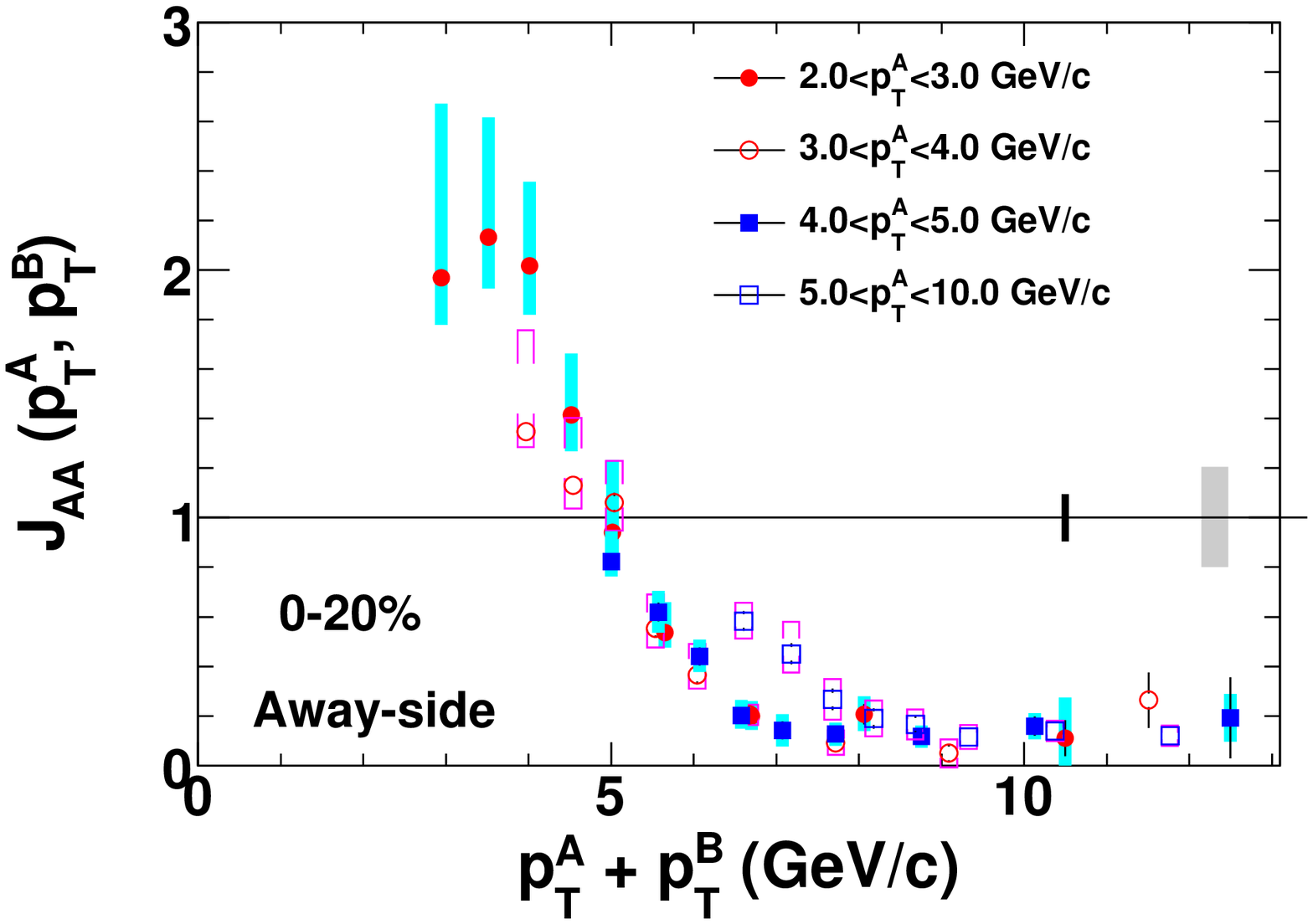,width=0.85\columnwidth}
\caption{\label{fig:5} The modification factor for hadron pair yield as function of $p_T^{sum} = p_T^A+p_T^B$ for the near-side (top) and away-side (bottom).
$p_T^{sum}$ condenses the 2-D correlation data in $p_T^A$ and $p_T^B$
space into a one dimensional plot. The STAR auto-correlation result~\cite{Daugherity:2008su} is divided by 3 (the lower end) to normalize the $\eta$ acceptance relative to PHENIX.}
\end{center}
\end{figure}

A physical observable better than PTY for describing the medium
response is the hadron pair yield JPY, the total number of
correlated pairs per-event, introduced in~\cite{Adare:2008cq}. The
modification of pair yield in Au+Au collision can be characterized
by $J_{AA}$:
\begin{eqnarray}
\label{eq:jaa} &&J_{\rm{AA}}(p_{\rm T}^{\rm a},p_{\rm T}^{\rm b},\Delta\phi) =
\frac{\rm{JPY}^{\rm{A+A}}}{\langle N_{\rm{coll}}\rangle\;
\rm{JPY}^{p+p}}\\\nonumber &&={\frac{1}{\sigma_{\rm
A+A}}\frac{d^3\sigma_{\rm
jet\_ind}^{\rm{A+A}}}{dp_{\rm T}^{\rm a}dp_{\rm T}^{\rm b}d\Delta\phi}} \mathord{\left/
 {\vphantom {\frac{d^3\sigma_{\rm{jet\_ind}}}{dp_{\rm T}^{\rm a}dp_{\rm T}^{\rm b}d\Delta\phi} }} \right.
 \kern-\nulldelimiterspace}
{\frac{\langle N_{\rm{coll}}\rangle}{\sigma_{ p+p}}\frac{d^3\sigma_{\rm
jet\_ind}^{p+p}}{dp_{\rm T}^{\rm a}dp_{\rm T}^{\rm b}d\Delta\phi}} \label{eq:6}
\end{eqnarray}
$J_{AA}$ quantify the medium modification of hadron pair yield from
the expected yield, in a way similar to $R_{AA}$ for describing the
modification of single hadron yield. The hadron pair yield is
proportional to the di-jet yield, and in the absence of nuclear
effects, it should scale with $N_{\rm coll}$, and $J_{AA}=1$.
Fig.~\ref{fig:5} shows $J_{AA}$ as a function of pair proxy energy
($p_T^{sum} = p_T^A+p_T^B$) for the near- (top panel) and away-side
(bottom panel). In contrast to a constant suppression at large
$p_T^{sum}$, the pair yields are less suppressed at $p_T^{sum}<6-8$
GeV/$c$. This reflects directly the energy transport that
redistributes energy of the quenched jets to low $p_T$ hadrons
(i.e. medium response). We would like to point out that $p_T^{sum}$
is a natural variable for the near-side correlation since it
approximates the original jet energy. In fact the data show an
approximate scaling in $p_T^{sum}$. Even the away-side data tend to
group together, because the medium response increases with
away-side jet energy which in turn increases with $p_T^{A,B}$.
%The STAR autocorrelation result~\cite{md} is shown as a single point at $2\langle p_T\rangle\sim1$ GeV/$c$.

\section{Consequence of Triggering on Medium Response.}
As argued above, the soft triggers (at $p_T<5$ GeV/c) have three
possible origins: jet fragmentation, hadrons that have no
correlation which leads to dilution of per-trigger yield, or
triggers come directly from medium response (e.g. the cone and the
ridge). In many theoretical models, the energy loss processes lead
to heating of the partons in the local medium. This local heating
effectively boosts bulk partons to higher energy, which then
hadronize into particles that are correlated with the quenched
jets. Since this type of mechanisms does not require new particle
production, they are more effective in generating large yield of
correlated pairs than those involving gluon radiation. However,
this contribution has a steeper slope in $p_T$ than the jet
fragmentation contribution. Thus it is important at $p_T<5$ GeV/c,
but become less important at higher $p_T$. In a simple jet
absorption picture, the yield of fragmentation hadrons from
survived jets scales with $R_{AA,o}$ (the constant suppression
factor at high $p_T$). The yield of medium response hadrons should
scale with the number of quenched jets, i.e. $1-R_{AA,o}$.

In general, hadrons in correlated pairs can be chosen from either
survived jets or quenched jets. Therefore, jet-induced pairs can be
divided into three groups. {\bf jet-jet pair}: both hadrons come
from fragmentation of survived jets. {\bf jet-medium pair}: one
hadron comes from fragmentation of survived jet, the second hadron
comes from medium feedback of quenched jet. {\bf medium-medium
pair}: both hadrons come from medium feedback of quenched jets. The
rates of their contributions to inter-jet pairs scale approximately
as $R_{AA,o}^2$, $R_{AA,o}(1-R_{AA,o})$ and $(1-R_{AA,o})^2$.
Clearly, in this simple picture, the stronger the suppression is,
the larger the contribution from medium-medium pairs is. In the
limit $R_{AA,o}\rightarrow0$, the medium-medium contribution
naturally dominates.

\begin{figure}
\begin{center}
\resizebox{1\linewidth}{!}{%
  \includegraphics{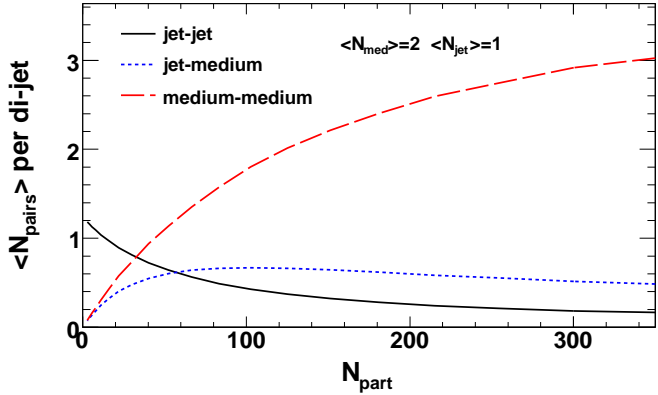}
}
\caption{\label{fig:m1} The centrality dependence of pair yield normalized by number of dijets
from jet-jet, jet-medium and medium-medium sources.}
\end{center}
\end{figure}

We performed a simple simulation based on jet absorption
picture~\cite{Drees:2003zh} to investigate contributions from the
three sources, focusing on the medium-medium pairs which were not
considered in many previous models. The details of the simulation
can be found in~\cite{Jia:2008vk}. The idea is to generate di-jets
according to collision density distribution in the overlap plane
($\rho_{ncoll} (x,y)$), and attenuate them as they traverse a
medium whose density is proportional to the participant nucleon
density ($\rho_{npart} (x,y)$)~\cite{Drees:2003zh}. Both
$\rho_{ncoll}$ and $\rho_{part}$ were generated via a Monte-Carlo
glauber model code. If a jet survives the medium, it is converted
into $N_{jet}$ hadron via fragmentation, otherwise it is converted
into $N_{med}$ medium response hadrons. $N_{jet}$ and $N_{med}$ are
assumed to have a poisson distribution with a mean of 1 and 2
respectively. The former corresponds to multiplicity of hadron in
1-4 GeV/c from a 6 GeV/c jet. The later is chosen to take into
account the observed enhancement in 1-4 GeV/c seen in the data.
Fig.~\ref{fig:m1} compares the yields for jet-jet, jet-medium and
medium-medium pairs as a function of centrality. The medium-medium
contribution increases with $N_{part}$ and dominates the yield in
central collisions.

To describe the $\Delta\phi$ distribution, we adjust the jet
fragmentation kinematics to match the $p+p$ correlation
data~\cite{Adler:2005ad}. We also assume that only 30\% of the
inter-jet pairs are detected, so as to account for finite detector
acceptance and the swing of the away-side jet in pseudo-rapidity.
The medium response hadrons are assumed to be emitted equally at
$\pi\pm D$ (D=1.1). This step makes sure that the shape of jet-jet
and jet-medium pairs, those included in traditional models, matches
the measured away-side distribution, so we can study the impact of
medium-medium pairs to the $\Delta\phi$ distribution.

\begin{figure}[th]
\resizebox{1\linewidth}{!}{%
  \includegraphics{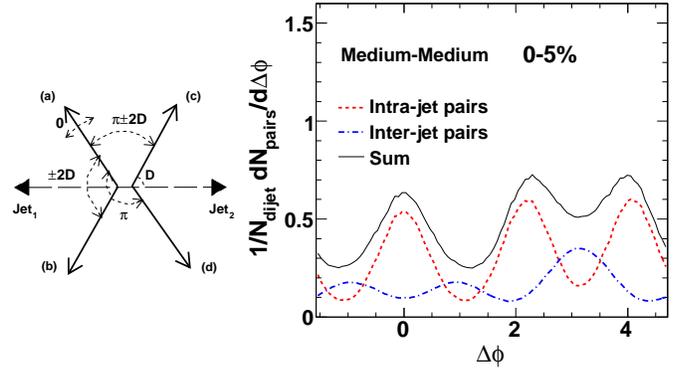}
}
\caption{\label{fig:m2}  $\Delta\phi$ distributions in 0-5\% central Au+Au collisions for medium-medium pairs. The diagram at the left
 illustrates the typical emission directions (solid arrows) and $\Delta\phi$ values (numbers).}
\end{figure}

The emission directions for hadrons contributing to medium-medium
pairs are illustrated by left panel of Fig.~\ref{fig:m2}. In this
picture, both jets are quenched and converted into hadrons at angle
$\pm D$ radians relative to the original jet direction. The pairs
constructed from these hadrons, when plotted in $\Delta\phi$,
should concentrate at $\Delta\phi\sim0$, $\pi$, $\pm 2D$ and
$\pi\pm2D$. The intra-jet pairs split up into three branches at
$\Delta\phi\sim0,2D$ and $-2D$, while the inter-jet pairs have a
sizeable peak at $\Delta\phi\sim\pi$ and two small satellite peaks
around $\Delta\phi\sim\pi\pm2D$. Since the value of $D=1.1$ implies
$2D \approx \pi-D$, the pairs from the same jets (at $\pm2D$)
coincide with the location of jet-medium pairs (at $\pi\pm D$).
Thus in our simulation, the medium-medium contribution alone
already accounts for the double-humped structure on the away-side.
Note that most of the pairs originate from the same jets. If there
are some mechanisms which broaden the hadrons from the same jet in
pseudo-rapidity, then both the near-side peak and away-side
shoulders should be elongated in $\Delta\eta$. Thus our model
provides a natural explanation for the similarities between the
ridge and the cone.

\begin{figure}[th]
\resizebox{1\linewidth}{!}{%
  \includegraphics{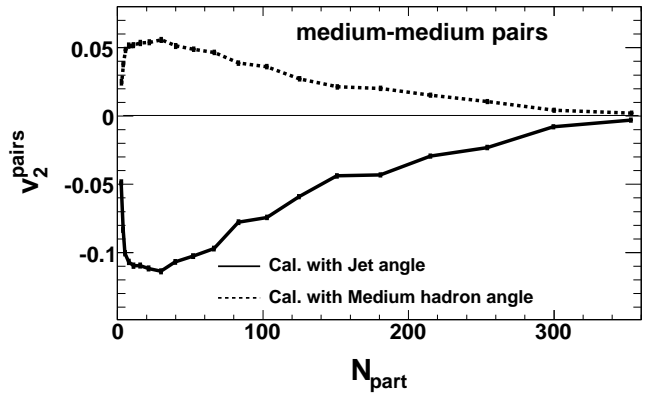}
}
\caption{\label{fig:m3} The $v_2$ for and medium-medium  pairs
calculated as $\langle cos2(\phi-\Psi_{RP})\rangle$, where $\phi$ is the azimuthal angle of jets (solid lines) or medium response hadrons (dotted lines). }
\end{figure}

%This is similar to the negative $v_2$ predicted for jet-conversion photons~\cite{Turbide:2005bz}.
Due to the path length difference of their parent partons, jet
absorption also leads to a positive anisotropy for the survived
jets, characterized by a positive $v_2$, $v_{2,sur}$. Since the
originally generated jets are isotropic, the $v_2$ of the quenched
jets should have an opposite sign relative to quenched jets:
$v_{2,quen}=-v_{2,sur}$. However, the medium response hadrons are
emitted at a large angle (D=1.1 radian) relative to the original
jet direction, which significantly dilutes the observed $v_2$
values for the medium response hadrons as shown by
Fig.~\ref{fig:m3}. So if we indeed are triggering on the medium
response hadrons, the pair $v_2$ should be smaller than that for
inclusive hadrons.

We note that the anisotropy studied in this model originates purely
from jet-quenching, other mechanisms could introduce new sources of
$v_2$ for hadron pairs, which can change the values presented in
Fig.~\ref{fig:m3}. A measurement of the reaction plane dependence
of the hadron pair yield (instead of the per-trigger yield), and
the corresponding $v_2$ parameter, should be useful in helping us
to constrain the underlying mechanisms of the medium response.
%Experimentally, the reaction plane dependence of the per-trigger yield has been measured. But a more useful measurement is the reaction plane dependence of the hadron pair yield, and the corresponding $v_2$ parameter for hadron pairs.

\begin{figure}[th]
\resizebox{1\linewidth}{!}{%
  \includegraphics{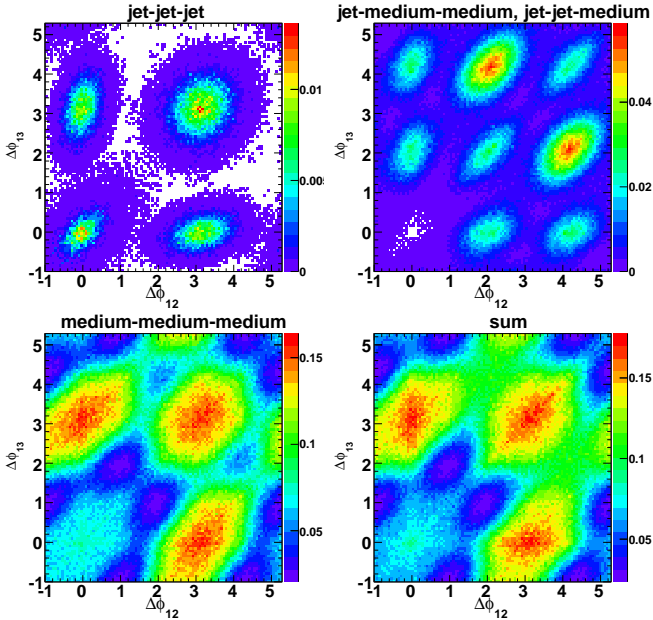}
}
\caption{\label{fig:m5}Azimuthal distributions for
triplets from various sources in 0-5\% Au+Au collisions.}
\end{figure}

We also studied the consequence of triggering on medium response
for three particle correlation. There are three types of triplets:
jet-jet-jet (j-j-j) triplets, jet -medium- medium (j-m-m), jet
-jet- medium (j-j-m) triplets and medium -medium- medium (m-m-m)
triplets. The standard coordinate
system~\cite{Ajitanand:2005xa,Abelev:2008nd} is adopted, i.e, we
plot the azimuth angle difference between particle 1 and 2
($\Delta\phi_{12}$) vs azimuth angle difference between particle 1
and 3 ($\Delta\phi_{13}$) as shown in Fig.~\ref{fig:m5}. The j-j-m
and j-m-m contributions lead to the conventional off-diagonal terms
which were used to identify the mach-cone pattern in the
data~\cite{Abelev:2008nd,Renk:2007rv}. The dominating contribution
comes from m-m-m term, which has a more complicated shape due to
the large deflection angle {\it w.r.t} the original jet direction.
The m-m-m term appears as several broad peaks in the
$\Delta\phi_{1,2}$ and $\Delta\phi_{1,3}$ plane. It's away-side
peak is so spread out that it shadows the modest off-diagonal
contribution from j-m-m and j-j-m terms. Clearly, the shape of the
three-particle correlation signal depends strongly on whether we
trigger on jet fragmentation (j-m-m, j-j-m) or medium response
(m-m-m), even though the correlation shape from medium-medium and
jet-medium can be similar. Our study shows that the interpretation
of the three particle correlation results are complicated if the
triggers come from the medium response.

\section{Additional Thoughts}
Most di-hadron correlation analyses rely on the following
two-source model formula to obtain the jet signal by subtracting
the background term~\cite{Adler:2005ee,Adare:2008cq}:
\begin{eqnarray}
C(\Delta\phi) = J(\Delta\phi) +\xi(1+2v_2^Av_2^B\cos2\Delta\phi)
\label{eq:3}
\end{eqnarray}
where $\xi$ is the background level, which is a number close to one
in central collisions. This approach implicitly assumes that either
there is no jet-flow cross-term, or the cross-term is included in
the definition of the jet function $J(\Delta\phi)$. Currently, the
main debate over Eq.~\ref{eq:3} centers on what $v_2$ values to use
for triggers ($v_2^A$) and partners ($v_2^B$). This debate is
caused by a proliferation of different flow measurements, designed
to remove non-flow bias and/or event-by-event fluctuation effects.
In our opinion, these additional sources contribute to
$C(\Delta\phi)$, but they should not be included in the jet
function $J(\Delta\phi)$. Instead, they should be included in the
flow background, i.e. the $v_2^A, v_2^B$ used in the background
subtraction should include various non-flow effects (other than
jet) and event by event fluctuations. If we do not include them in
the flow background, then we would need to subtract their
contributions later from the jet yield.~\footnote{The only caveat
is that the non-flow and event by event contributions are assumed
to have $\cos2\Delta\phi$ shape. Based on how the $C(\Delta\phi)$
varies with angle {\it w.r.t} reaction plane~\cite{Jia:2005ab}, it
is a fairly good assumption.}

\begin{figure*}[!]
\begin{center}
\resizebox{0.8\linewidth}{!}{%
  \includegraphics{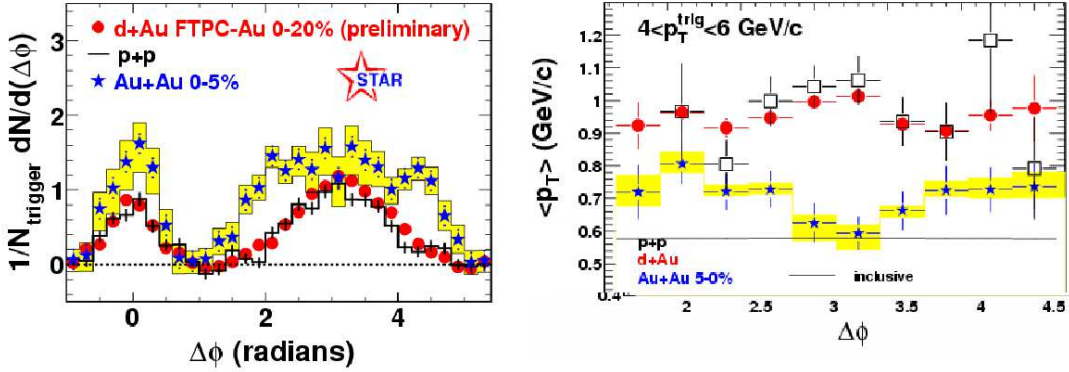}
} \caption{\label{fig:7} Correlation for $4-6 \otimes 0.15-4$ GeV/c
from STAR collaboration in $d$+Au and 0-5\% Au+Au
collisions~\cite{Wang:2005cg}: a) the per-trigger yield
distribution in $\Delta\phi$, b) the mean $p_T$ of partners as
function of $\Delta\phi$. }
\end{center}
\end{figure*}

The $v_2$ values used in the background subtraction are different
between STAR and PHENIX. PHENIX collaboration uses reaction plane
$v_2$ based on forward detectors. STAR collaboration use the
average of reaction plane $v_2$ and four particle culminant
$v_2$~\cite{Adams:2005ph}. The later is about 20\% smaller than the
former, and the difference between the two is quote as the
systematic error. As a result, the $v_2$ used by STAR is about 10\%
smaller than that for PHENIX, but their differences is covered by
the large $v_2$ error quoted by STAR. But because of this
difference, STAR collaboration does not see a very pronounced dip
in the $dN/\Delta\phi$ distribution as in PHENIX
(Fig.\ref{fig:7}a). However, they show that if one chose a
different quantity, such as the mean $p_T$ which characterizes the
energy flow, one observes a clear dip at $\Delta\phi \sim\pi$
(Fig.\ref{fig:7}b). This shows that away-side shape also depends on
the physical observable. Existence of a dip in multiplicity
distribution is not a sufficient condition for Mach cone, because
it could be filled up by the punch-through jet, or the jet wake.

\begin{figure*}[!]
\begin{center}
\resizebox{0.7\linewidth}{!}{%
  \includegraphics{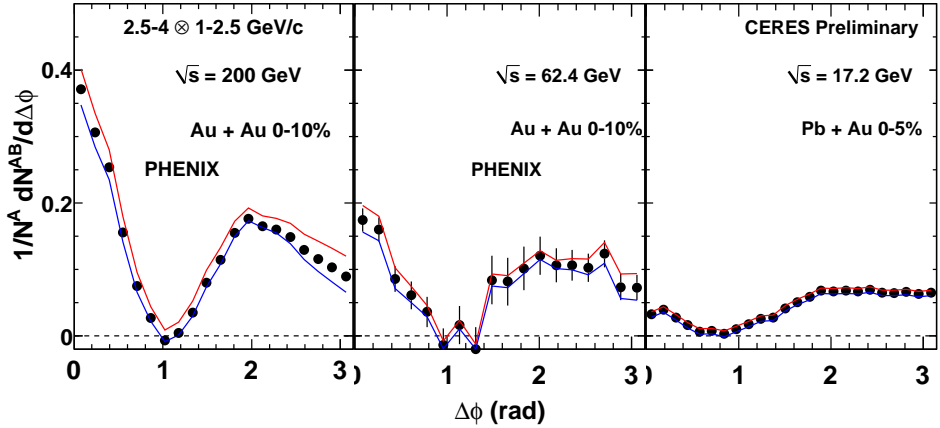}
} \caption{\label{fig:8} The $\Delta\phi$ distribution in central
collisions for three collision energies in central collisions. }
\end{center}
\end{figure*}

Important insights on the origin of the cone and the ridge have
been provided by a energy scan from SPS to RHIC. A strong
modification of the away-side correlation was observed at the top
SPS energy ($\sqrt{s_{NN}}=17.2$ GeV)~\cite{Ploskon:2007es}. There,
the strong away-side broadening has been used to argue for a
similar interpretation (such as Mach cone) as for results at RHIC.
However a quantitative analysis of the energy dependence of the
modification patterns (see Fig.\ref{fig:8}) shows that the yield of
medium response are quite different between RHIC and SPS energies.
The near-side yield drop by almost factor of 8 going from 200 GeV
to 17 GeV, the away-side shoulder yield drops by a factor of 2 in
the same energy range, while little dependence of the yield on
$\sqrt{s}$ is observed for the away-side head region, where the jet
fragmentation is expected to be important. This suggests a much
weaker medium response at SPS energy (the ridge almost disappeared
and cone strongly suppressed) than that at RHIC, which probably
implies different mechanisms are in play at SPS energy, such as a
stronger Cronin effect combined with a weaker energy loss at lower
energy~\cite{adare:2008cx}.

\section{Conclusion}

First several years of RHIC data on jet-induced particle
correlations have revealed a fascinating and rich set of insights
on the mechanisms for jet medium interactions. The $p_T$, PID and
centrality dependence of the correlation patterns are observed to
be consistent with competition between jet fragmentation and medium
response. Fragmentation of survived jets dominates the correlation
signal at high $p_T$, while the medium response to quenched jets
dominates the correlation signal at low $p_T$. It is also observed
that the medium response at the near-side (the ridge) and the
away-side (the cone) share similar properties.

Understanding the medium response requires knowledge of the origins
of the hadrons in the pairs at $p_T<5$ GeV/c. Both the spectra and
correlation results suggest we may be triggering on medium response
at intermediate $p_T$. A simple jet absorption model is employed to
investigate the roles of pairs from medium response (medium-medium
pair) on the pair shape and pair yield. The correlations among
medium response particles were shown to dominate the pair yield at
intermediate $p_T$, while preserving the di-hadron correlation
patterns in $\Delta\phi$. These medium-medium pairs are expected to
have significantly reduced anisotropy due to the smearing caused by
the large emission angle {\it w.r.t} the original jet direction.
The model suggests a common mechanistic origin for the ridge and
the cone, and provides a natural explanation for their
similarities.

We also touched on the flow background subtraction in the
two-source model framework. We argue that the non-flow effects and
event by event fluctuation effects should not be taken out of the
$v_2$ used in the subtraction. We also show that the shape of the
away-side $\Delta\phi$ distribution shows a more significant dip in
mean $p_T$ than in multiplicity. Finally the energy scan data
suggests that the medium response mechanisms might be different at
SPS from those at RHIC energies.

% BibTeX users please use
% \bibliographystyle{}
% \bibliography{}

\end{document}